\documentclass[aps,prb,twocolumn,showpacs]{revtex4-1}
\usepackage{graphicx,psfrag}
\usepackage{amsmath}
\usepackage{amsfonts}
\usepackage{amssymb}
\usepackage{bm}
\usepackage{color}

\def\bea{\begin{eqnarray}}
\def\eea{\end{eqnarray}}
\def\be{\begin{equation}}
\def\ee{\end{equation}}
\def\bpm{\begin{pmatrix}}
\def\epm{\end{pmatrix}}

\def\sign{\mathop{\rm sign}}

\newcommand{\ve}{\varepsilon}

\newcommand{\D}{\Delta}

\newcommand{\bk}{{\bm k}}

\newcommand{\bS}{{\bm S}}
\newcommand{\bs}{{\bm s}}
\newcommand{\bp}{{\bm p}}
\newcommand{\bq}{{\bm q}}

\newcommand{\br}{{\bm r}}

\newcommand{\ba}{{\bm a}}

\newcommand{\bv}{{\bm v}}
\newcommand{\bde}{{\bm e}}
\newcommand{\bsigma}{{\bm \sigma}}

\newcommand{\barf}{{\bar f}}
\newcommand{\barc}{{\bar c}}

\newcommand{\1}{\mathbb{1}}

\newcommand{\up}{{\uparrow}}
\newcommand{\down}{{\downarrow}}
\newcommand{\ti}{\text{imp}}
\newcommand{\calGamma} {\mathcal{I}}  
\newcommand{\Do} {\Delta_1}  

\begin{document}

\title{Overscreened Kondo fixed point in $S=1$ spin liquid}
\author{Maksym Serbyn, T. Senthil,  and Patrick A. Lee}
\affiliation{Department of Physics, Massachusetts Institute of
Technology, Cambridge, Massachusetts 02139}
\date{\today}

\begin{abstract}
We propose a possible realization of the overscreened Kondo impurity problem by a magnetic $s=1/2$ impurity embedded in a two-dimensional $S=1$ $U(1)$ spin liquid with a Fermi surface. This problem contains an interesting interplay between non-Fermi-liquid behavior induced by a $U(1)$ gauge field coupled to fermions and  a non-Fermi-liquid fixed point in the overscreened Kondo problem. Using a large-$N$ expansion together with an expansion in the dynamical exponent of the gauge field, we find that the coupling to the gauge field leads to weak but observable changes in the physical properties of the system at the overscreened Kondo fixed point. We discuss the extrapolation of this result to a physical case and argue that the realization of overscreened Kondo physics could lead to observations of effects due to gauge fields. 
\end{abstract}


\maketitle

\section{Introduction}

Impurity models constitute an important chapter in modern condensed matter physics. Since the original paper by Kondo~\cite{Kondo} considering electron sea screening a single impurity spin, this problem has attracted significant theoretical and experimental attention.~\cite{Abrikosov,AbrikosovMigdal,NozieresBlandin,ReadNewns,AffleckNPB90,*Affleck91,*AffleckNPB91,ParcolletPRL,ParcolletPRB,Sengupta,Fabrizio-PRB96,GanPRL,*Gan,Gogolin,Gordon,Hewson,Cox,Shlotmann-rev,KondoBeteReview} %
More recently, impurity physics has been studied in the context of strongly interacting systems. Numerous examples include~\cite{VojtaRev} an impurity in systems with vanishing density of states~\cite{WithoffPRL90,FradkinPRB96,*VojtaPRB04,*VojtaPRB04-2,*FritzPRB04,*FritzPRB06}  %
high temperature superconductors,~\cite{KhaliulinPRB97,*NagaosaPRB95,*NagaosaPRL97,*PepinPRL98,*WangLeePRL02} %
and quantum magnets.~\cite{SachdevSc99,*SachdevPRB03,*HoglundPRL07,*HoglundPRL07-2,*MetlitskiPRB08,FlorensPRL06,*FlorensPRB07,KolezhukPRB06,KimJCMP2008,DhochakPRL10,RibeiroLee} %
Quantum magnets are particularly versatile as a host system, having a large number of possible ground states with different low energy excitations. 

In this paper we consider a spin-half impurity embedded in a spin-1 quantum paramagnet with a spin liquid ground state.~\cite{Anderson1,*Anderson2}  We consider the situation where the low energy excitations of the paramagnet are described by emergent fermionic excitations with a Fermi surface, coupled to a $U(1)$ gauge field. This study is motivated by the recent appearance of several $S=1$ spin liquid candidate materials.~\cite{Nakatsuji,Goodenough} Theoretically, a number of  spin liquid ground states for spin-1 system have been proposed.~\cite{Senthil-NiGaS,TrebstPRB09,Tsunetsugu,Ng-short,*Ng-long,GroverPRL11,SerbynS1,XuPRL12,Samuel} %
One possible scenario involves emergence of three fermionic excitations carrying spin-1 quantum numbers.~\cite{Ng-short,SerbynS1,Samuel} Assuming that Fermi surfaces of these excitations are not destroyed by a pairing instability, we obtain the host system that is considered below.

Impurity physics in a spin-$1/2$ spin liquids has been considered in the context of bosonic spin liquids,~\cite{FlorensPRL06} algebraic spin liquids,~\cite{KolezhukPRB06,KimJCMP2008} and spin liquids with a Fermi surface.~\cite{RibeiroLee} In particular, Ribeiro and one of us in Ref.~\onlinecite{RibeiroLee} concluded that physics of a spin-$1/2$ impurity embedded in a spin liquid with $S=1/2$ fermionic excitations is similar to that of the conventional Kondo problem.~\cite{Hewson} In what follows we argue that a spin-$1/2$ impurity in a $S=1$ spin liquid with a Fermi surface realizes overscreened Kondo physics. Although our results are qualitatively similar to the overscreened Kondo effect in conventional systems, there are observable differences due to the presence of an emergent gauge field coupled to spinons. 

Our findings suggest that an impurity in a $S=1$ spin liquid can be used to probe fermionic excitations. As these  excitations do not carry charge, their experimental detection is a difficult problem. Different experimental probes have been suggested in the context of spin-$1/2$ spin liquids.~\cite{MotrunichPRB06,NormanPRL09,Wing-HoRaman,WingHoRIXS,MrossPRB84,ZhouLee} We suggest that the realization of overscreened Kondo physics is a possible way to unravel physics of spin-one spin liquid, allowing probes of fermionic excitations, as well as the presence of an emergent gauge field.

Overscreened Kondo physics is realized in  multichannel Kondo models, where a single spin is coupled to $N$ copies (flavors) of itinerant electrons.~\cite{NozieresBlandin}  On the one hand, such a generalization of original Kondo  model may be seen as merely a theoretical tool, allowing a perturbative expansion in $1/N$. On the other hand, the physics changes drastically depending on the interrelation between impurity spin length, $s$, and the number of flavors coupled to the impurity. When the number of flavors, $N$, is just enough or less than needed to screen the impurity spin, $N\leq 2s$, antiferromagnetic coupling between the impurity and electrons flows to infinity in the infrared, meaning that at low temperatures impurity spin is screened to the maximum possible extent by electrons, resulting in Fermi-liquid behavior.~\cite{ReadNewns,Hewson} However, in the overscreened regime, $N>2s$, i.e. when there are more channels than required to screen the impurity spin, the system has a non-Fermi-liquid fixed point.~\cite{Affleck91,GanPRL,ParcolletPRL,Sengupta} This state is characterized by singularities in different physical observables, such as impurity spin susceptibility, specific heat, \emph{etc.} It  is particularly interesting as a solvable example of a system with a non-Fermi-liquid fixed point.~\cite{Affleck91}  Despite the rich and interesting physics, the overscreened regime of Kondo model has only few realizations~(in particular quantum dots and two levels systems.~\cite{Cox,Shlotmann-rev,Gordon}). Hence our system is also interesting as a possible implementation of overscreened Kondo physics.

Qualitatively, the problem of a spin-half impurity hosted by isotropic $S=1$ spin liquid looks similar to the conventional overscreened Kondo impurity model. When coupled antiferromagneticaly, itinerant excitations carrying spin-1 quantum numbers cannot screen the impurity. However,  the presence of a gauge field effectively enforcing a single occupancy constraint for fermionic excitations, makes these two problems different. Even without the impurity, fermions are in a non-Fermi-liquid regime~\cite{LeeNagaosa_PRB46,Altshuler94,Altshuler95,KimLee,Motrunich_PRB72,SungSik,Mross} due to the gauge field. The fermion propagator is dressed by a singular self-energy, so there are no well defined quasiparticle excitations in the system.  This is manifested, for example, in the singular behavior of the specific heat $C\propto T^{2/3}$ in two dimensions at low temperatures.~\cite{LeeNagaosa_PRB46,Motrunich_PRB72}

Coupling the impurity to fermions with non-Fermi-liquid behavior allows us to study the interplay between the gauge field induced non-Fermi-liquid behavior and the Kondo non-Fermi-liquid fixed point. The conventional approach to the Kondo problem is either an exact solution by mapping it onto one-dimensional problem,~\cite{AffleckNPB90,KondoBeteReview} or $1/N$ expansion. Both methods are not directly applicable in our case. The presence of gauge field impedes the mapping of our model to a one dimensional problem in the radial channel.  On the other hand, a rigorous $1/N$ expansion is not possible, due to singular self-energy corrections~\cite{SungSik, Metlitski1}. The latter issue has been recently resolved in the paper by Mross~\emph{et. al.},~\cite{Mross} where a controlled double expansion scheme has been provided. It combines the $1/N$ expansion with an expansion in another small parameter (related to the dynamical critical exponent of the gauge field).

We adopt the recently developed double expansion method~\cite{Mross} to our problem. Since the double expansion includes the large $N$ limit, we expect to have a perturbatively accessible fixed point. At leading order, the gauge field does not affect the position of this non-Fermi-liquid Kondo fixed point. However, it leads to corrections to the scaling dimension of the Kondo coupling. Assuming that the results obtained using the double expansion interpolate to the physical case, we conclude that physical properties such as impurity spin susceptibility, specific heat, etc. are still characterized by singular behavior. Unlike the case of the Kondo model in the regime of perfect screening,~\cite{RibeiroLee} where the coupling to the gauge field has no consequences to leading order in $1/N$,  in our case the gauge field influences Kondo physics.

The rest of the paper is organized as follows. In the remainder of this Section, we introduce the basics of our model, diagram technique and briefly explain the idea behind double expansion. In Section~\ref{Sec:FixedPoint} we first review known results for the $\beta$-function in the overscreened Kondo problem without the gauge field. Afterwards, we calculate the $\beta$-function with the gauge field and study the changes in scaling behavior of different physical quantities. Finally, in Section~\ref{Sec:Discussion} we discuss the extrapolation of our findings beyond the double expansion, and comment on possible experimental realizations and experiments to detect Kondo physics. Details regarding the calculation of corrections to the $\beta$-function due to gauge field are given in the Appendix.

\subsection{Spin liquid with fermionic excitations and impurity \label{Sec:IntroducingModel}}
Our starting point is a spin Hamiltonian on a lattice consisting of spin-1 sites,
\begin{equation} \label{Eq:Hspin}
  H_\text{spin}
  =
  \sum_{ij} \left[
  J_{ij}\bS_i\cdot \bS_j 
  +
  K_{ij}(\bS_i\cdot \bS_j)^2 
  \right]
  +
  \ldots,
\end{equation}
where ellipses denote other possible terms needed for stabilizing the $U(1)$ spin liquid phase with a Fermi surface. We do not specify the lattice, since only an effective low energy description of a spin liquid phase is relevant in what follows. However we note that recent work~\cite{Samuel} shows  evidence for the possibility of stabilizing such a phase on a triangular lattice with nearest neighbors bilinear, and biquadratic spin interactions along with ring exchange terms. In the large-$U$ limit of a half-filled Hubbard model one can easily derive the coupling between a given lattice spin at site $i$ and an impurity spin in a form
\begin{equation} \label{Eq:Himp}
  H_\text{imp}
  =
  J_K \bS_i\cdot \bs_\ti.
\end{equation} 
This procedure leads to $J_K>0$ corresponding to antiferromagnetic coupling. 

We assume that Hamiltonian in Eq.~(\ref{Eq:Hspin}) has a spin liquid ground state with fermionic excitations and a Fermi surface. The low energy description of such a state is a theory of fermions strongly coupled to a $U(1)$ gauge field. Referring the reader to the literature for more detailed discussions,~\cite{Wenbook} we only summarize results. A spin-1 operator at a given site is represented using \emph{three} fermion operators, ${\tilde f}_{\lambda}$, $\lambda=1,2,3$ as
\begin{equation} \label{Eq:spinons}
  \bS_i
  =
  \sum_{\lambda,\rho=1}^3{\tilde f}^\dagger_{i\lambda} {\bm I}^{\lambda\rho} {\tilde f}_{i\rho},
\end{equation}
with ${\bm I}^{\lambda\rho}$ being the set of three spin-1 matrices~(generators of $SU(2)$ in spin-1 representation).~\cite{Ng-short} In order to remove unphysical states from the Hilbert space, introduced by the representation in Eq.~(\ref{Eq:spinons}), one has to enforce a single occupancy constraint on each site.  Fermionic  ${\tilde f}_{\lambda}$ are the low energy excitations of the spin liquid, carrying spin-1 quantum numbers. In addition, the low energy description contains a $U(1)$ gauge field, coupled to fermions ${\tilde f}_{i\lambda}$ and enforces the single occupancy constraint. 

Before proceeding further, let us reiterate the question of interest. We want to understand if the non-Fermi-liquid fixed point of a conventional overscreened Kondo model is changed by the presence of the gauge field in the bulk. The model outlined above provides us with a particular setup to study the influence of the non-Fermi-liquid bulk on the overscreened Kondo fixed point. However, in order to have  control over calculations we need to resort to the large-$N$ limit. The crucial requirement for the generalization procedure is to retain the presence of the overscreened Kondo fixed point. We choose a model with $N$ species of \emph{spin-half} fermions, $f_{i \alpha m}$ with $\alpha=\up,\down$, and $m=1\ldots N$ as a large-$N$ generalization. This is the simplest model which allows for controllable calculations. 

The corresponding Lagrangian for our generalized model may be split into a fermionic part~(including coupling to gauge field and impurity spin), and a gauge field Lagrangian,
\begin{equation} \label{Eq:Ltot}
  L
  =
  L_\text{fermion}
  +
  L_\text{gauge}.
\end{equation}
The generalized fermion Lagrangian becomes:
\begin{multline} \label{Eq:Ltriplon}
  L_\text{fermion}
  =
  \int d\tau \sum_{\bk,m,\alpha}
  \Big(
  \barf_{\bk \alpha m} (\partial_\tau - \ve_\bk) f_{\bk \alpha m }
  \\
  -
  \frac{e}{\sqrt{N}} \sum_\bq  f^\dagger_{\bk+\frac{\bq}{2} \alpha m} \bv(\bk) \cdot \ba_{\bq}  f_{\bk-\frac{\bq}{2} \alpha m}
  -\frac{J_K}{N}  \bS(0)\cdot \bs 
  \Big)
  ,
\end{multline}
where we use imaginary time. In accordance with the discussion above, fermion operators $f_{i\alpha m}$ now carry spin-1/2 quantum numbers~($\alpha=\up,\down$). We omit the time component of gauge field from the coupling, since it is screened,~\cite{KimLee} and do not write the diamagnetic term, including it in the gauge field propagator. The fermion spin at $\br =0$ is
\begin{equation} \label{Eq:S0}
  \bS(0) 
  = 
  \frac{1}{{\cal N}}\sum_{\bk,\bp,\alpha,\beta,m} f^\dagger_{\bk\alpha m } \frac{\bsigma^{\alpha\beta}}{2} f_{\bp \beta m},
\end{equation}
with $\bsigma = (\sigma^x,\sigma^y,\sigma^z)$ being the set of three Pauli matrices, and ${\cal N}$ being a number of sites in the lattice.  In what follows, Greek indices label spin projection, $\alpha,\beta,\ldots = \up, \down$, whereas Latin indices  $m,n,\ldots=1\ldots N$ label channels.  

The gauge field Lagrangian is
\begin{equation}\label{Eq:Lgauge}
  L_\text{gauge}
  =
 \frac12 \int \frac{d \bq\, d\omega}{(2\pi^3)} 
 a^*_i(\bq,\omega) D^{-1}_{ij}(\bq,i\omega) a_j(\bq,\omega)
 ,
\end{equation}
where the time component  of the gauge field is excluded. The bare gauge field propagator is zero, since the gauge field is not dynamical but rather represents fluctuations around the mean field ansatz.  However, nontrivial dynamics are generated if one accounts for coupling to fermions, leading to a non-zero $D^{-1}_{ij}(\bq,i\omega)$,  discussed in Section~\ref{Sec:DoubleExpansion}.

\subsection{Diagram technique \label{Sec:DiagramTechnique}}
\begin{figure}
\includegraphics[width=0.85\columnwidth]{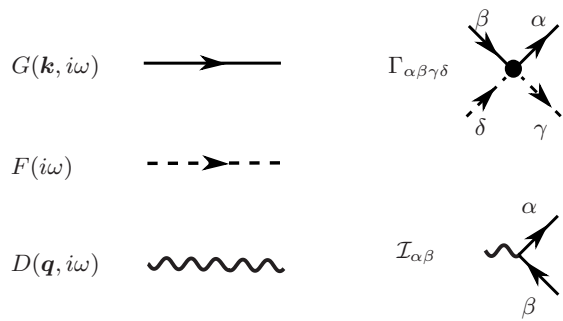}
\caption{Summary of rules for diagram technique. Solid, dashed and wavy lines represent fermion, pseudofermion and gauge field propagator respectively. Also, interaction vertices of fermions with gauge field, $\calGamma_{\alpha\beta}$, and fermions with impurity pseudofermions, $\Gamma_{\alpha\beta\gamma\delta}$, are shown.  All objects are diagonal in flavor indices, which are thus suppressed.  \label{Fig:DiagBasics}}
\end{figure}
The impurity spin is conveniently represented via fermionic operators,
\begin{equation}\label{Eq:Simp}
  \bs
  =
  \sum_{\alpha,\beta} c^\dagger_\alpha \frac{\bsigma^{\alpha\beta}}{2} c_\beta,
\end{equation}
where $c^\dagger_{\up,\down}$ ($c_{\up,\down}$) are creation (annihilation) operators of spin up or down pseudofermions.~\cite{Abrikosov} In what follows we use the term ``pseudofermions'' to distinguish the operators $c_\alpha$ from the operators $f_{\bk \alpha m}$, which describe low energy excitations in the spin liquid. A faithful representation of spin via fermion operators requires an additional constraint to exclude doubly occupied and empty states from the Hilbert space. However, in the case of a single spin-$1/2$ operator, $\bs$, in Eq.~(\ref{Eq:Simp}), gives zero when acting on unphysical states in the Hilbert space. Therefore, one can ignore the constraint in this case,~\cite{Abrikosov} writing the impurity Lagrangian as
\begin{equation} \label{Eq:Limp}
  L_\text{imp}
  =
  \int d\tau\,
  \barc_m (\partial_\tau - \mu_\text{imp}) c_m,
\end{equation}
where $\mu_\text{imp} = +0$ is the chemical potential for impurity pseudofermions. 

The rules for diagram technique, following from Eqs.~(\ref{Eq:Ltriplon})-(\ref{Eq:Limp}) are summarized in Fig.~\ref{Fig:DiagBasics}. Propagators for fermions and pseudofermions along with interaction vertices are given by
\begin{subequations}\label{Eq:DiagramRules}
\begin{eqnarray} \label{Eq:Gfun}
  G_{\alpha \beta}^{mn}(\bk, i\omega)
  &=&
  \frac{\delta_{\alpha\beta}\delta_{mn}}{i\omega - \xi_\bk -  \Sigma(i\omega)},
\\ \label{Eq:Ffun}
  F_{\alpha\beta}( i\omega)
  &=&
  \frac{\delta_{\alpha\beta}}{i\omega - 0},
  \\ \label{Eq:Gvert}
  \Gamma^{mn}_{\alpha\beta\gamma\delta}
  &=& 
  - \frac{J_K}{4} \bsigma_{\alpha\beta}\cdot \bsigma_{\gamma\delta} \delta_{mn},
\\ \label{Eq:CalGvert}
  \calGamma^{mn}_{\alpha\beta }
  &=&
   -\frac{e}{\sqrt{N}} \bv_\bk \delta_{\alpha\beta}\delta_{mn},
\end{eqnarray}
\end{subequations}
where $\xi_\bk = \ve_\bk - \mu$ is the fermion energy relative to the Fermi surface. The self-energy, included in fermion Greens function~[Eq.~(\ref{Eq:Gfun})]  is discussed below. We note that interaction between fermions and the impurity is local in real space. Therefore, in Fourier space, the momenta of two fermion operators in the impurity interaction vertex~[Eq.~(\ref{Eq:Gvert})] are unrelated. Fermion propagator and interaction vertices are diagonal in flavor indices. Thus the only contribution of flavor indices is an extra factor $N$ for every loop of fermions, and they will be suppressed in the remainder of the paper.  

\subsection{Double expansion \label{Sec:DoubleExpansion}}
\begin{figure}
 \includegraphics[width=0.85\columnwidth]{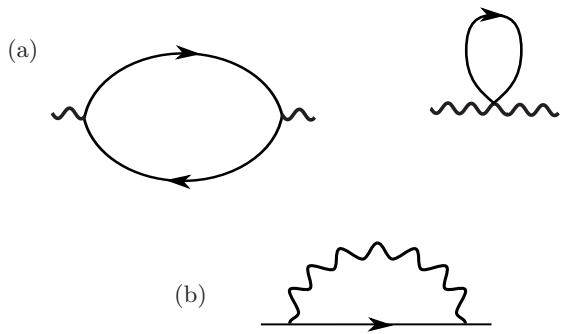}
\caption{\label{Fig:gauge} (a) Self-energy of the gauge field due to interaction with fermions. Second diagram describes diamagnetic contribution. (b) Self-energy of fermions due to interactions with gauge field in the leading order in $1/N$.}
\end{figure}

We briefly review the  double expansion framework introduced in Ref.~\onlinecite{Mross}. First, we specify dynamically generated propagator of the gauge field. To leading order, the propagator is given by the fermion bubble with current vertices along with diamagnetic term shown in Fig.~\ref{Fig:gauge}~(a). In the Coulomb gauge, $\nabla\cdot \ba = 0$, the propagator is transverse and can be written as~\cite{KimLee,Altshuler94,Mross} 
\begin{subequations} \label{Eq:Dfun}
\begin{eqnarray} \label{Eq:Dgen}
  D^{-1}_{ij}(\bq,i\omega) 
  =
  \left(\delta_{ij}-\frac{q_iq_j}{q^2}\right) D^{-1}_0(\bq,\omega),
  \\ \label{Eq:D0}
  D^{-1}_{0}(\bq,i\omega) 
  =
  \gamma \frac{|\omega|}{q} 
  +
  \chi_0 q^{z_b-1},
\end{eqnarray}
\end{subequations}
with $z_b=3$ and $\gamma = 2n/k_F$, $\chi_0 = 1/(24\pi m)$ for fermions with quadratic dispersion. Note that the Landau damping term is non-zero only for $|\omega|<v_F q$.  We assume $z_b$ takes general value $z_b<3$ and use it as a control parameter. This approach is consistent because terms $q^{z_b-1}$ for $z_b<3$ are non-local. Since $z_b$ is not going to be renormalized within perturbation theory, it is a valid control parameter.  

The singular form of the  gauge propagator~[Eq.~(\ref{Eq:Dfun})] leads to a singular self-energy correction for fermions.  In the leading order in $1/N$, the diagram in Fig.~\ref{Fig:gauge} (b) gives us:~\cite{KimLee, SungSik, Mross}
\begin{subequations}
\label{Eq:Sigma-Lambda}
\begin{eqnarray} \label{Eq:Sigma}
  \Sigma (i\omega)
  =
  -i\lambda\, |\omega|^{2/z_b}\sign \omega,
  \\ \label{Eq:lambda}
  \lambda 
  =
  \frac{e^2}{N} \frac{v_F}{\gamma} \frac{1}{4\pi \sin\frac{2\pi}{z_b}}\left(\frac{\gamma}{\chi_0}\right)^{2/z_b}.
\end{eqnarray}
\end{subequations}
For $z_b>2$, the self-energy is more important than the bare $i\omega$ term in Greens function~[Eq.~(\ref{Eq:Gfun})],  when $|\omega|<\omega_0$.  The energy scale $\omega_0$ is set by a combination parameters $\gamma$, $\chi_0$ and  $v_F$~[see Eq.~(\ref{Eq:omega0})] and is of order of Fermi energy, the only energy scale related to fermions. 

When the self-energy, Eq.~(\ref{Eq:Sigma-Lambda}), is singular compared to the bare frequency dependence of the fermions' Greens function, a factor of $1/N$ in the fermion self-energy leads to an extra power of $N$ in the \emph{numerator} of the Greens function.  This spoils naive power counting in the $1/N$ expansion~\cite{Metlitski1, SungSik} requiring a summation of an infinite series of diagrams of a particular topology (genus) at any given order in $1/N$.  However, if we assume a gauge field dynamical exponent,~\footnote{In notations adopted in Ref.~\onlinecite{KimLee}, $\eta= z_b-1 = 1 +\ve$} $z_b=2+\ve$, and take the double scaling limit:~\cite{Mross} 
\begin{equation}\label{Eq:DoubleScaling}
  \ve \rightarrow 0,
  \qquad
  N\rightarrow \infty,
  \qquad
  \ve N = \mathrm{const}, 
\end{equation} 
we obtain finite $\lambda \propto 1/(N\ve)$ in Eq.~(\ref{Eq:lambda}),  rather than $\lambda\propto 1/N \to 0$. The absence of the factor $1/N$ in front of the self-energy restores naive power counting, where the gauge field interaction vertex contributes $1/\sqrt{N}$ and each fermion loop gives a factor of $N$.

Finally, before proceeding further, we rewrite $\Sigma(i\omega)$ in a simplified form, valid in the double scaling limit,
\begin{equation} \label{Eq:SigmaDouble}
   \Sigma (i\omega)
  =
  -i \frac{1}{N\ve}
  \omega \left|\omega_0\over\omega\right|^{\ve/2},
\end{equation}
where scale $\omega_0$ is explicitly given by
\begin{equation} \label{Eq:omega0}
  \omega_0
  =
  \frac{\chi_0}{\gamma}\left(\frac{e^2}{2\pi^2}\frac{v_F}{\chi_0}\right)^{2/\ve}.
\end{equation}

\section{Perturbatively accessible fixed point \label{Sec:FixedPoint}}

The renormalization group~(RG) approach in conjunction with $1/N$ expansion has been proven to be fruitful when applied to the conventional Kondo impurity problem~\cite{Abrikosov,AbrikosovMigdal,Nozieres,NozieresBlandin,GanPRL,Gan}. The renormalization procedure is defined with respect to the fermion bandwidth, $D$. Eliminating states far away from the Fermi surface, one studies the induced flow in the dimensionless coupling $g = \nu J_K$~($\nu$ is the density of states assumed to be constant within the whole band). The corresponding $\beta$-function is defined as
\begin{equation} \label{Eq:beta-def}
  \beta(g)
  =
  \frac{d\log g }{d\log D},
\end{equation}
and can be calculated perturbatively in $g$. This simplification is brought by the $1/N$ expansion and is justified in vicinity of fixed point located at small $g^*\propto O(1/N)$.

When there is a gauge field coupled to fermions, the RG approach still can be applied. However, it  requires some modifications. The usual $1/N$ expansion has to be replaced  by the double expansion discussed above, but the RG flow is still defined with respect to bandwidth, $D$. Coupling of the fermions to the gauge field, $e^2$, is treated as a constant, since a single impurity can not change its flow under RG. Likewise in the conventional Kondo problem, there exists a perturbatively accessible fixed point. After briefly reviewing the calculation for RG flow in conventional Kondo problem, we calculate the $\beta$-function in the presence of the gauge field and obtain physical properties in the vicinity of the fixed point.

\subsection{$\beta$-function in conventional Kondo problem  \label{Sec:betafunction0}}
\begin{figure}
\includegraphics[width=0.95\columnwidth]{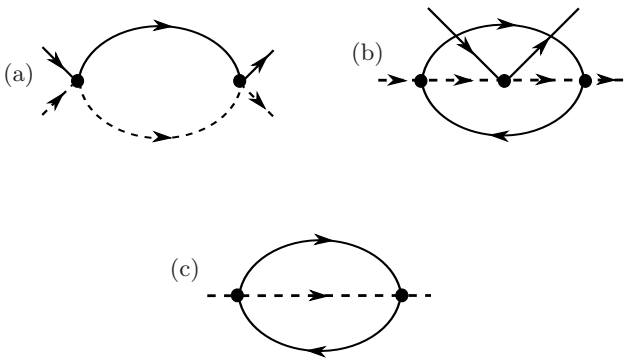}
\caption{\label{Fig:Diagbeta} Diagrams contributing to the $\beta$-function in the leading order in $1/N$. Diagrams (a) and (b) describe corrections to the vertex in the second and third orders of perturbation theory~(symmetric counterpart of diagram (a) with direction of one of the fermion line changed is not shown). Diagram (c) is the correction to the  self-energy of pseudofermions,  contributing to $\beta$-function via renormalization of $Z$-factor.}
\end{figure}

While reviewing the RG procedure for the usual Kondo impurity problem we mostly follow Refs.~\onlinecite{Abrikosov,AbrikosovMigdal}. Renormalization of the dimensionless coupling $g$ in the leading order is given by diagrams shown in Fig.~\ref{Fig:Diagbeta}. Diagrams (a) and (b) in Fig.~\ref{Fig:Diagbeta} represent corrections to the bare interaction vertex in the second and third orders of perturbation theory. These are the only diagrams up to the third order, which are logarithmically divergent and thus renormalize the coupling. Note, that  diagrams (a) and (b) describe the contributions of  the same order, since the latter diagram in addition to extra power of $g\propto 1/N$ has a factor of $N$ from the fermion loop. Diagram (c) in Fig.~\ref{Fig:Diagbeta} describes renormalization of $Z$-factor of pseudofermions and also contributes to the $\beta$-function.  

Calculation of the diagrams in Fig.~\ref{Fig:Diagbeta} gives the $\beta$-function:
\begin{equation} \label{Eq:oldbeta}
  \beta(g)
  =
  \frac{d\log g }{d\log D}
  =
  g^2 - \frac{N}{2}g^3 +\ldots,
\end{equation}
where ellipses denote terms $O(1/N^3)$ coming from higher order diagrams. These terms have the form  $C_1 N g^4 + C_2 N^2 g^5$ with coefficients $C_{1,2}$ readily available in the literature~\cite{AndreiPRL84,Affleck91,GanPRL, Gan} and listed in Appendix~\ref{App:DiagramCalc}. 
One can easily solve for a non-trivial unstable fixed point of this $\beta$-function:
\begin{eqnarray} \label{Eq:g*0}
  g^*
  &=&
  \frac{2}{N} + \ldots,
  \\ \label{Eq:Delta0}
  \Delta_0
  &=&
  -\beta'(g^*)
  =
  \frac{2}{N}+\ldots,
\end{eqnarray}
where $\Delta_0$ is the negative slope of the $\beta$ function at this fixed point. Ellipses here stand for terms $O(1/N^2)$. We see that $g^*$ is indeed small in $1/N$, justifying the use of perturbation theory.

\subsection{Correction to $\beta$-function due to gauge field\label{Sec:betafunction}}
\begin{figure}
\includegraphics[width=0.95\columnwidth]{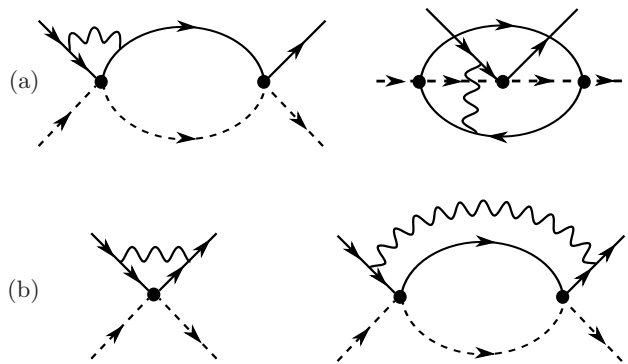}
\caption{Two types of vertex corrections in the leading order in $1/N$ due to gauge field. (a) Example of vanishing diagrams with a single gauge propagator connected by at least one end to the internal line. (b) Non-vanishing corrections, representing a new non-local vertex (first diagram) and example of diagram leading to its renormalization (second diagram).  \label{Fig:DiagVertex}}
\end{figure}
As we shall see, within the double expansion framework, the gauge field produces a small correction to the regular $\beta$-function. Therefore, it suffices to consider the effect of the gauge field to leading order. 

There are two types of effects related to the gauge field. First, the gauge field destroys well-defined quasiparticle, leading to non-Fermi-liquid behavior. This is manifested by the  singular self-energy due to the gauge field in the fermion propagator~[Eq.~(\ref{Eq:SigmaDouble}]. Therefore, one has to recalculate diagrams in Fig.~\ref{Fig:Diagbeta} using the fermion propagator which contains the self-energy.  A lengthy but straightforward calculation (see Appendix~\ref{App:DiagramCalc} for details) yields an answer identical to the case without gauge field, however, with a modified divergent logarithm. Namely, the standard $\log$-divergent contributions are replaced by  
\begin{equation} \label{Eq:logkappa}
  \log \frac{D}{\omega} \to \log\frac{D}{\omega^{1-\kappa} \omega_0^\kappa},
  \qquad 
  \kappa
  =
 \frac{1}{2N}\frac{1}{1+(N\ve)^{-1}},
\end{equation}
where energy scale $\omega_0 \propto D$ was defined in Eq.~(\ref{Eq:omega0}).

Another effect of the gauge field is the appearance of vertex corrections. All diagrams describing vertex corrections can be split into two classes with representatives of each class depicted in Fig.~\ref{Fig:DiagVertex}~(a) and (b) respectively. Diagrams belonging to the first class have at least one of the ends of the gauge field propagator connected to the internal fermion line. In Appendix~\ref{App:VertexCorr} we show that due to the transverse character of the gauge field propagator and the locality of interaction with the impurity, all diagrams of this type with single gauge propagator exactly vanish.~\cite{RibeiroLee}   

In the vertex correction diagrams attributed to the second class, the gauge field propagator connects two external lines. One can think about these diagrams as describing a new interaction vertex~[first diagram in Fig.~\ref{Fig:DiagVertex}~(b)] and its renormalization~[all other diagrams of this type]. This new vertex contains an additional small factor $1/N$, compared to the original impurity interaction vertex. However, it is non-local, since it depends strongly on the relation between outgoing and incoming fermion momenta, $\bk$ and $\bp$ respectively. The vertex is logarithmically divergent when the transferred momentum $|\bk-\bp|$ is close to $2 k_F$ and is small otherwise.~\cite{KimLee,Mross,Altshuler94} The flow of this vertex to leading order in $1/N$ is identical to the flow of the standard vertex. Thus it does not influence the scaling in the vicinity of the fixed point. The effect of this vertex is to provide subleading corrections to different observables~(due to the extra factor $1/N$). Therefore, in what follows we do not consider this vertex.

As we demonstrated, no new diagrams contribute to the $\beta$-function up to order $1/N^3$. Calculation of diagrams in Fig.~\ref{Fig:Diagbeta} with self-energy included into fermionic propagator gives us
\begin{equation} \label{Eq:betanew}
  \beta(g)
  =
  (1-\kappa) \left(g^2-\frac{N}{2}g^3\right) 
  +\ldots, 
\end{equation}
where ellipses represent terms $O(1/N^3)$.  These terms have to be included for consistency, since $\kappa\propto 1/N$, but are identical to those in the $\beta$-function without gauge field, Eq.~(\ref{Eq:oldbeta})~[account for gauge field in these terms will produce corrections $O(1/N^4)$]. 

The obtained $\beta$-function, Eq.~(\ref{Eq:betanew}), differs from the $\beta$-function without gauge field, Eq.~(\ref{Eq:oldbeta}), by terms $O(1/N^3)$. This correction does not shift the fixed point $g^*$ even at order $1/N^2$, compared to fixed point Eq.~(\ref{Eq:g*0}). However, it modifies the slope of $\beta(g)$ at the fixed point,
\begin{equation} \label{Eq:delta}
  \Do 
  =
  \frac{2}{N}\left(1-\kappa\right) + \ldots 
  =
   (1-\kappa) \Delta_0,
\end{equation}
compared to the slope for the case of the conventional Kondo problem, $\Delta_0$~[Eq.~(\ref{Eq:Delta0})].  Below, the slope of the $\beta$-function, $\Delta_1$  will be used to determine the flow of the coupling in the vicinity of the fixed point, as well as the singular behavior of different measurable quantities.~\cite{Gan,GanPRL} Therefore, the difference between $\Delta_0$ and $\Delta$ modifies the behavior of different observables compared to conventional Kondo problem.
\subsection{Observables \label{Sec:Observables}}

In order to understand how the non-Fermi-liquid fixed point manifests itself in observables, we first find the dependence of the running coupling constant~$g_R(\omega)$ on~$\omega$. It can be determined from the flow equation $dg_R(\omega)/d\log\omega = -\beta(g_R)$, by employing results for $\beta$-function and its slope at the fixed point. Denoting the bare value of coupling at $\omega=D$, as $g_R(D)=g$, we have:~\cite{GanPRL,Gan}
\begin{equation} \label{Eq:gR}
  g_R(\omega)
  =
  g^* - \zeta \left(\frac{\omega}{T_K}\right)^\Delta,
\end{equation}
where $\Delta$ is the slope of the $\beta$-function which depends on the presence of the coupling to the gauge field. The position of the fixed point, $g^*$, Eq.~(\ref{Eq:g*0}) is not influenced by the gauge field. The Kondo temperature is $T_K=D g^{N/2} \exp(-1/g)$ and $\zeta = (g^*)^{1+N\Delta/2}\exp(-\Delta/g^*)$. We assumed here that $\omega<T_K$ and that the initial value of coupling, $g$, is small. 

The power law behavior of the running coupling leads to a similar behavior in different physical quantities. Corrections to different measurable properties within perturbation theory can be expressed as a series in coupling $g$. Applying the renormalization group to this series results in singular behavior as a function of frequency or temperature with exponent proportional to $\Delta$. For the case when there is no gauge field present, this program has been implemented in Refs.~\onlinecite{GanPRL, Gan}. Generalization of this procedure to the case with the gauge field is straightforward. 

The main effect of the gauge field is always related to the different values of slope of $\beta$-function, $\Delta$. Without a gauge field $\D=\D_0$ is given by Eq.~(\ref{Eq:Delta0}). When there is a gauge field, we have $\D=\Do$, specified in Eq.~(\ref{Eq:delta}). While for thermodynamical quantities this is the only effect, transport properties and other quantities acquire small corrections to prefactors which are not given here. 

Calculating the contribution of impurity to the  imaginary part of  self-energy of fermions gives the scattering rate due to the impurity.   As a function of frequency, it acquires a cusp at $\omega=0$, $\nu \tau^{-1}(\omega) \propto 1 - N\zeta (\omega/T_K)^\Delta$  (cf. with a Lorentzian shape for a Fermi liquid fixed point). The correction to the resistivity due to Kondo interaction has a similar form, however, it is of little interest due to neutral character of fermionic excitations in spin liquid. The correction to the heat conductivity is potentially more interesting, 
\begin{equation} \label{Eq:kappa_th}
  \frac{\delta \kappa_{\text{th}}}{\kappa_{\text{th}}^0}
  \propto  
   n_\ti  \left[1 - N\zeta \left(\frac{\omega}{T_K}\right)^\Delta\right],
\end{equation}
as it can be distinguished using its proportionality to the impurity concentration, $n_\ti$.   

Also one can calculate corrections to different thermodynamic properties. A rigorous calculations of the self-energy allows us to find impurity specific heat with a critical exponent $\alpha=2\Delta$:
\begin{equation} \label{Eq:Cimp}
  C_\ti
  =
  \frac{3\pi^2}{2}\zeta^2\Delta \left(\frac{T}{T_K}\right)^{2\Delta}.
\end{equation}
 Magnetic properties, such as the impurity susceptibility as temperature $T\to 0$ and dependence of magnetization on the field $h=\mu_B H$ at $T= 0$  are given by
\begin{eqnarray} \label{Eq:chiimp}
  \chi_\ti
  &=&
  \left(\frac{N\zeta}{2}\right)^2 \frac{1}{T} \left(\frac{T}{T_K}\right)^{2\Delta},
  \\\label{Eq:Mimp}
  M
  &=&
  \frac{N\zeta}{2}  \left(\frac{h}{T_K}\right)^{\Delta}.
\end{eqnarray}
Likewise, it is possible to find an expression for fermion-fermion, fermion-impurity and impurity-impurity susceptibilities~\cite{GanPRL}. Lastly, we list results for $\chi_\ti''(\omega,T )/\omega$ which is a contribution to the NMR relaxation rate due to the impurity. Its behavior is again specified by $\Delta$, and for~$\omega\ll T$ 
\begin{equation} \label{Eq:NMR}
   \frac{\chi_\ti''(\omega,T )}{\omega}  \propto 
   T^{2\Delta-2}. 
\end{equation}

\section{Discussion \label{Sec:Discussion}}

We have investigated the effect of the gauge field strongly coupled to fermions at a non-Fermi liquid overscreened Kondo fixed point. Using the double expansion framework, we demonstrated that the gauge field  does not alter the position of the perturbatively accessible non-Fermi-liquid fixed point, but leads to corrections to exponents characterizing the behavior of different physical properties in the vicinity of the fixed point. In particular, it ``softens'' the non-analytic behavior of specific heat, magnetization, spin susceptibility, compared  to those for a Kondo problem without the gauge field. The physical origin of this effect is the ``smearing'' of the sharp quasiparticles by the gauge field.

Let us discuss the extrapolation to the physical case. In order to have a control over calculations, we worked in the double expansion limit, Eq.~(\ref{Eq:DoubleScaling}), with $N$ species of spin-half fermions. We note, that if the coupling to gauge field was absent, the considered model for $N=4$ corresponds to the one channel of spin-one itinerant moments coupled to impurity.~\cite{Sengupta,Gogolin} The same equivalence was checked to hold in our perturbative calculations of $\beta$-function for the case when there is a coupling to the gauge field. 

Thus, we expect that the physical case corresponds to $N=4$, $\ve = 1$. Assuming that our results can be extrapolated to these values of $N$ and  $\ve$, we can argue that the non-Fermi-liquid Kondo fixed point is not destroyed by the presence of a gauge field. However, we expect singularities in different physical properties related to the non-Fermi-liquid fixed point to be weakened compared to their values without gauge field. In such a case, the realization of overscreened Kondo physics in $S=1$ spin liquid may be used not only to observe neutral fermionic excitations, but as evidence for the presence of a gauge field. Indeed, non-Fermi-liquid behavior may be used as an indication of fermionic excitations present in the system. At the same time, the  difference of observed scalings from those for the case without a gauge field~\cite{Affleck91,Gogolin} may be used as a litmus test for the presence of a gauge field coupled to fermions.  From an experimental point of view, specific heat (proportional to impurity concentration), as well as spin susceptibility and NRM relaxation rate are the most promising probes.

It is instructive to compare the role of the gauge field in our case to the case of the Kondo model in the regime of perfect screening, Ref.~\onlinecite{RibeiroLee}. In the latter case, the system flows to the infinite coupling fixed point, and the results of Ref.~\onlinecite{RibeiroLee} show no changes in impurity specific heat and spin susceptibility due to the presence of the gauge field. 

Finally, we discuss possible experimental realizations of our proposal. In a recent experiments~\cite{Nakatsuji,Goodenough} materials that could possibly realize the spin liquid with fermionic excitations~\cite{Ng-short,SerbynS1,Samuel} has been found. One can speculate on the possible stabilization of $U(1)$ spin liquid phase in the same or similar type of materials. The presence of spin-half impurities in such a phase would realize the scenario considered in our work. Another way to implement the discussed physics is to go to lower dimensions. A gapless phase for spin-1 with bilinear and biquadratic interaction has been established for a certain range of couplings.~\cite{FathPRB44,*FathPRB47,ItoiPRB97,Affleck_JPA01,LauchliPRB74}. A spin-half impurity in such a chain is expected to realize overscreened Kondo physics. A detailed consideration of this problem will be presented elsewhere. 

M.S. acknowledges discussions with David Mross and John McGreevy. This work was supported by NSF DMR 1005434~(T.~S.) and NSF DMR 1104498~(P.~A.~L.).

\appendix

\section{Calculation of diagrams for $\beta$-function \label{App:DiagramCalc}}

In this Appendix we present calculation of diagrams  in Fig.~\ref{Fig:Diagbeta} with fermion propagator, Eq.~(\ref{Eq:Gfun}), containing self-energy due to gauge field.  Detailed calculation of these diagrams without gauge field can be found, for example, in Refs.~\onlinecite{Abrikosov,AbrikosovMigdal}. 

First we consider diagram in Fig.~\ref{Fig:Diagbeta}~(a), describing second order correction to the dimensionless coupling   $g=\nu J_K$. We will be interested only in the  logarithmically divergent part of the diagram.  Using zero temperature Matsubara diagram technique and implying summation over repeated indices  we can write for the correction to impurity interaction vertex: 
\begin{multline} \label{Eq:Gamma-a1}
  \Gamma^{(a1)}_{\alpha\beta\gamma\delta}
  =
 \left(\frac{J_K}{4}\right)^2 (\bsigma_{\alpha \rho}\cdot \bsigma_{\gamma \lambda})(\bsigma_{\rho \beta}\cdot \bsigma_{\lambda\delta})
  \\
  \times
  \int \frac{d\bk\, d\omega_1}{(2\pi)^3} F(-i\omega_1)G(\bk,i\omega+i\omega_1).
\end{multline}
Symmetric counterpart of diagram (a) with flipped direction of the propagation of the pseudofermions (not shown in Fig.~\ref{Fig:Diagbeta}) gives us,
\begin{multline} \label{Eq:Gamma-a2}
  \Gamma^{(a2)}_{\alpha\beta\gamma\delta}  = 
 \left(\frac{J_K}{4}\right)^2 (\bsigma_{\alpha \rho}\cdot \bsigma_{\lambda\delta})(\bsigma_{\rho\beta}\cdot \bsigma_{\gamma \lambda})
  \\
  \times
  \int \frac{d\bk\, d\omega_1}{(2\pi)^3} F(i\omega_1)G(\bk,i\omega+i\omega_1).
\end{multline}
After integrating over $\omega_1$ and changing integration variable from $\bk$ to $\xi = \ve_\bk-\mu$, we have similar expressions for  both diagrams: 
\begin{multline} \label{Eq:Gamma-a1-1}
  \Gamma^{(a1,2)}_{\alpha\beta\gamma\delta}
  =
  -
  \left(\frac{J_K}{4}\right)^2 (\mp 2 \bsigma_{\alpha\beta}\cdot \bsigma_{\gamma\delta} + 3 \delta_{\alpha\beta}\delta_{\gamma\delta})
  \\
  \times
  \nu \int d\xi \frac{\theta(\pm \xi)}{i\omega (1+\frac{1}{N\ve}\left|\frac{\omega_0}{\omega}\right|^{\ve/2})-\xi },
\end{multline}
We perform integration over $\xi$, retaining only logarithmical part.  Collecting results for both diagrams and going to real frequency domain, we get:
\begin{multline} \label{Eq:Gamma-a12-2}
  \Gamma^{(a)}_{\alpha\beta\gamma\delta}
  =
  -4\nu \left(\frac{J_K}{4}\right)^2 (\bsigma_{\alpha\beta}\cdot \bsigma_{\gamma\delta})
  \\
  \times \log \frac{D}{|\omega| (1+\frac{1}{N\ve}\left|\frac{\omega_0}{\omega}\right|^{\ve/2})}.
\end{multline}
We expand logarithm in $\ve$ to the leading order and collect both terms into single logarithm again:
\begin{multline} \label{Eq:Log-expand}
  \log \frac{D}{|\omega| (1+\frac{1}{N\ve}\left|\frac{\omega_0}{\omega}\right|^{\ve/2})}
  =
  \log\frac{D}{|\omega|} - \kappa \log\frac{\omega_0}{|\omega|}
  \\
  =
  \log \frac{D}{|\omega|^{1-\kappa}\omega_0^\kappa}
\end{multline}
where $\kappa$ is small, $\kappa \propto O(1/N)$,
\begin{equation} \label{Eq:kappa-app}
 \kappa 
 =
 \frac{1}{2N} \frac{1}{1+\frac{1}{N\ve}}.
\end{equation}
Finally, we have
\begin{equation} \label{Eq:Gamma-a12-answ}
   \Gamma^{(a)}_{\alpha\beta\gamma\delta}
  =
  g    \log \frac{D}{|\omega|^{1-\kappa}\omega_0^\kappa}   
  \Gamma_{\alpha\beta\gamma\delta},
\end{equation}
where bare vertex $  \Gamma_{\alpha\beta\gamma\delta}$ is defined in Eq.~(\ref{Eq:Gvert}), and we retained only logarithmically divergent terms. Alternatively, we could expand in Greens function in $1/N$ already in Eqs.~(\ref{Eq:Gamma-a1})-(\ref{Eq:Gamma-a2}), reproducing the same result.

Calculations of vertex and $Z$-factor renormalization, described correspondingly by diagrams (b) and (c) in  Fig.~\ref{Fig:Diagbeta} are very similar. Indeed, in order to get impurity pseudofermions $Z$-factor, $Z_\ti$, we have to differentiate self-energy over $\omega$, what may be thought of as an introduction of additional vertex with zero incoming frequency. Therefore, below we present only details on the calculation for the derivative of self-energy and list the result for the vertex renormalization. 

Correction to the impurity self-energy described by diagram Fig.~\ref{Fig:Diagbeta}~(c)  is written as
\begin{multline} \label{Eq:Sigma-c-1}
  \Sigma^\ti(i\omega)
  =
  -6 N  \left(\frac{J_K}{4}\right)^2
  \int \frac{d\bk_1\, d\omega_1}{(2\pi)^3} \frac{d\bk_2\, d\Omega_2}{(2\pi)^3}
  \\ \times
   G(\bk_1,i\omega_1)G(\bk_2,i\omega_1+i\Omega_2)F(i\omega+i\Omega_2)
  ,
\end{multline}
where we omitted spin indices of external pseudofermions and associated $\delta$-function.
Renormalization of  $Z_\ti$ is given by the derivative of self-energy,
\begin{equation} \label{Eq:deltaZ-def}
  \delta Z_\ti
  =
  -\frac{\partial \Sigma^\ti(i\omega)}{\partial (i\omega)}.
\end{equation}
Integrating  over $\Omega_2$ in Eq.~(\ref{Eq:Sigma-c-1}), we have
\begin{equation} \label{Eq:deltaZ-2}
    \delta Z_\ti
  =
 - 6 N  \left(\frac{J_K}{4}\right)^2\frac{\partial}{\partial (i\omega)}
  \int \frac{d\bk_1\, d\bk_2}{(2\pi)^4} {\cal I}_{\bk_1,\bk_2,i\omega},
\end{equation}
\begin{equation} \label{Eq:calI-def}
 {\cal I}_{\bk_1,\bk_2,i\omega}
 = 
  \int \frac{d\omega_1}{2\pi}G(\bk_1,i\omega_1)G(\bk_2,i\omega_1-i\omega) \theta(-\xi_{\bk_2})
  .
\end{equation}
To simplify further calculations, we  expand in $\ve$ and $1/N$. Self-energy, Eq.~(\ref{Eq:SigmaDouble}), expanded to the leading order in~$\ve$ becomes:
\begin{equation} \label{Eq:SigmaDouble-exp}
  \Sigma(i\omega)
   =
  -i\omega \left(\frac{1}{N\ve}+\frac{1}{2N}\log \left|\frac{\omega_0}{\omega}\right|\right).
\end{equation}
Inserting this into fermion Greens function, Eq.~(\ref{Eq:Gfun}), and expanding  in $1/N$ we get:
\begin{multline} \label{Eq:Gexpansion}
 G(\bk_1,i\omega_1)
 =\tilde G(\bk_1,i\omega_1)
 \\
 -\frac{1}{2N}i\omega_1\log \left|\frac{\omega_0}{\omega_1}\right|[\tilde G(\bk_1,i\omega_1)]^2,
\end{multline}
where $\tilde G(\bk_1,i\omega_1)$ is defined as:
\begin{equation} \label{Eq:}
 \tilde G(\bk_1,i\omega_1)
 =\frac{1}{i\omega_1(1+\frac{1}{N\ve})-\xi_{\bk_1}}.
\end{equation}
Finally, expansion of the product of Greens function in $ {\cal I}_{\bk_1,\bk_2,i\omega}$, Eq.~(\ref{Eq:calI-def}), gives us:
\begin{multline} \label{Eq:calI-exp}
   {\cal I}_{\bk_1,\bk_2,i\omega}
=
  \int \frac{d\omega_1}{2\pi} \theta(-\xi_{\bk_2})
 \tilde G(\bk_1,i\omega_1)\tilde G(\bk_2,i\omega_1-i\omega)
 \\
 \times \left[1 
 -\frac{1}{2N}i\omega_1\log \left|\frac{\omega_0}{\omega_1}\right| \tilde G(\bk_1,i\omega_1)
 \right.
 \\
 \left.
 -\frac{1}{2N}i(\omega_1-\omega)\log \left|\frac{\omega_0}{\omega_1-\omega}\right| \tilde G(\bk_2,i\omega_1-i\omega) 
 \right].
\end{multline}
After integration over $\omega_1$,  zeroth order term in~(\ref{Eq:calI-exp}) yields
\begin{equation} \label{Eq:Ical-1}
   {\cal I}^{(0)}_{\bk_1,\bk_2,i\omega}
   =
  -\frac{\theta(\xi_{\bk_1})\theta(-\xi_{\bk_2})}{(1+\frac{1}{N\ve})[\xi_{\bk_1}+|\xi_{\bk_2}|-i(1+\frac{1}{N\ve})\omega]}.
\end{equation}
This is inserted into Eq.~(\ref{Eq:deltaZ-2}). After integration over momenta extra factors $(1+\frac{1}{N\ve})$ drop out and we reproduce the answer for the case without gauge field, $\delta Z_\ti^{(0)}  =   3N g^2/8 \log (D/|\omega|)$. 

Frequency integration for terms proportional to $1/N$ in Eq.~(\ref{Eq:calI-exp}) results into cumbersome expression. However, after integrations over $\xi_{\bk_1}$ and $\xi_{\bk_2}$ and extracting $\log$-divergent part we obtain $\delta Z_\ti^{(1)}  =  -3\kappa N g^2/8   \log (\omega_0/|\omega|)$, where $\kappa$ is defined in Eq.~(\ref{Eq:kappa-app}). Combining $\delta Z_\ti^{(0)}$ and $\delta Z_\ti^{(1)}$, we have for impurity pseudofermions $Z$-factor:
\begin{equation} \label{Eq:deltaZ-answ}
  Z_\ti
  =
  1+ \delta Z_\ti
  =
  1+\frac{3}{8}Ng^2  \log \frac{D}{|\omega|^{1-\kappa}\omega_0^\kappa} .
\end{equation}

Correction to the impurity interaction vertex, diagram Fig.~\ref{Fig:Diagbeta}~(b) is calculated in a similar way. Resulting contribution to the interaction vertex is 
\begin{equation} \label{Eq:Gamma-b-answ}
 \Gamma^{(b)}_{\alpha\beta\gamma\delta}
  =
  - \frac{N}{8} g^2    \log \frac{D}{|\omega|^{1-\kappa}\omega_0^\kappa}   \Gamma_{\alpha\beta\gamma\delta}
.
\end{equation}

Finally, renormalized coupling is 
\begin{equation} \label{Eq:gRen}
  g_R
  =
  \frac{g+\delta g}{Z_\ti},  
\end{equation}
where $Z_\ti$ is given by Eq.~(\ref{Eq:deltaZ-answ}), and $\delta g$ can be read from Eqs.~(\ref{Eq:Gamma-a12-answ}) and~(\ref{Eq:Gamma-b-answ}):
\begin{equation} \label{Eq:deltag}
 \frac{ \delta g}{g}
  =
  \left(g- \frac{N}{8} g^2 \right)  \log \frac{D}{|\omega|^{1-\kappa}\omega_0^\kappa} . 
\end{equation}
Using that $\omega_0\propto D$, we obtain the $\beta$-function:
\begin{multline} \label{Eq:beta-full}
  \beta(g)
  =
  \frac{d\log g }{d\log D}
  =
    (1-\kappa) \left(g^2-\frac{N}{2}g^3\right)  
    \\
  -  \frac{N}{2}(1+\log 2) g^4 + \frac{N^2}{4}g^5,
\end{multline}
where in the second line we included subleading terms $O(1/N^3)$ obtained in Ref.~\onlinecite{GanPRL, Gan}. Note, that Eq.~(\ref{Eq:beta-full}) is exact to the order $1/N^3$: corrections to subleading terms from to the gauge field  are of order $O(1/N^4)$ and thus can be ignored.

\section{Vertex corrections \label{App:VertexCorr}}
\begin{figure}
\includegraphics[width=0.45\columnwidth]{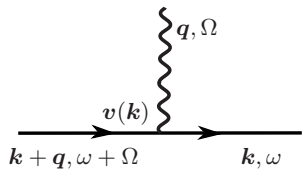}
\caption{Part of diagram with vertex corrections that makes the diagram vanish. \label{Fig:VertexPart}}
\end{figure}
In this Appendix we demonstrate that a subset of vertex corrections where gauge field propagator is connected to internal fermion Greens function vanish. Two examples of such diagrams are shown in Fig.~\ref{Fig:DiagVertex}~(a). It suffices to consider a part present in all diagrams, consisting of two Greens functions and a single gauge field vertex, Fig.~\ref{Fig:VertexPart}. Using notations adopted in Fig.~\ref{Fig:VertexPart}, we can write for the integral over momentum $\bk$ \begin{equation} \label{Eq:VertexZero}
  \int dk_x\, dk_y\, v_y(\bk) G(\bk, i\omega) G(\bk+q \bde_x, i\omega+i\Omega),
\end{equation}
where we assumed that $\bq$ has only $x$-component,  $\bq\parallel \bde_x$ and used fact that gauge field is transverse. Note that integration over $\bk$ does not involve any other functions due to the fact that interaction with impurity is local. It is integration over $k_y$ in Eq.~(\ref{Eq:VertexZero}) that makes the expression to be zero. Indeed, prefactor  $v_y(\bk)$ is odd under inversion of $k_y$, whereas both Greens functions do not change under $k_y\to -k_y$. Since this part is present in all diagrams in Fig.~\ref{Fig:DiagVertex} (b), all these diagrams vanish.

\end{document}